\documentclass[runningheads]{llncs}
\usepackage[T1]{fontenc}
\usepackage{graphicx}
\usepackage{xcolor}
\usepackage{siunitx}
\usepackage{amsmath}
\usepackage{amssymb}
\usepackage{subcaption}  % サブキャプションを使用するためのパッケージ
\usepackage{overpic}
\usepackage{color}
\usepackage{booktabs}
\usepackage{siunitx}

\allowdisplaybreaks

\begin{document}
\title{Comparative Analysis of Black-Box Optimization Methods for Weather Intervention Design}
\titlerunning{Black-Box Optimization toward Weather Intervention}

\author{Yuta Higuchi\inst{1}\orcidID{0009-0005-0728-5429} \and
Rikuto Nagai\inst{2}\orcidID{0009-0002-2498-9370}\and\newline
Atsushi Okazaki\inst{3}\orcidID{0000-0002-4598-0589} \and
Masaki Ogura\inst{1}\orcidID{0000-0002-3857-3942} \and\newline
Naoki Wakamiya\inst{2}\orcidID{0000-0002-6195-6087}}

\authorrunning{Higuchi, Y., Nagai, R., Okazaki, A., Ogura, M., Wakamiya, N.}
%
% 1 [所属先名], [住所], [都市名] [郵便番号], [国名]
% 2 [所属先名], [住所], [都市名] [郵便番号], [国名]
% [メールアドレス]
% [ウェブサイトのURL]
\institute{Hiroshima University, 1-3-2 Kagamiyama, Higashi-hiroshima, Hiroshima 739-8511, Japan \\
%{\{b212693,oguram\}の, の後に跡にspace NG
\email{\{b212693,oguram\}@hiroshima-u.ac.jp}\and
Osaka University,
  1-5 Yamadaoka,
Suita, Osaka 565-0871, Japan\\
\email{\{rk-nagai,wakamiya\}@ist.osaka-u.ac.jp}\and
Chiba University, 1-33 Yayoi, Inage, Chiba, Chiba 263-8522, Japan\\
\email{atsushi.okazaki@chiba-u.jp}}

\maketitle            
\begin{abstract}
As climate change increases the threat of weather-related disasters, research on weather control is gaining importance. The objective of weather control is to mitigate disaster risks by administering interventions with optimal timing, location, and intensity. However, the optimization process is highly challenging due to the vast scale and complexity of weather phenomena, which introduces two major challenges. First, obtaining accurate gradient information for optimization is difficult. In addition, numerical weather prediction (NWP) models demand enormous computational resources, necessitating parameter optimization with minimal function evaluations.
To address these challenges, this study proposes a method for designing weather interventions based on black-box optimization, which enables efficient exploration without requiring gradient information. The proposed method is evaluated in two distinct control scenarios: one-shot initial value intervention and sequential intervention based on model predictive control. Furthermore, a comparative analysis is conducted among four representative black-box optimization methods in terms of total rainfall reduction. 
Experimental results show that Bayesian optimization achieves higher control effectiveness than the others, particularly in high-dimensional search spaces. These findings suggest that Bayesian optimization is a highly effective approach for weather intervention computation.

\end{abstract}
\section{Introduction}
As global warming progresses, weather-related disasters such as hurricanes, floods, and torrential rain have become increasingly frequent and severe across many regions of the world.
Over the 35 years leading up to 2014, the number of weather-related loss events approximately tripled, with total economic losses reaching up to US\$125 billion in 2005~\cite{hoeppe2016}.
Various studies on weather control have been conducted under this background~\cite{hoffman2002,dong2021}.
However, because weather control involves complex and large-scale meteorological phenomena, it faces the following three major challenges: (1) identifying effective interventions, (2) selecting feasible interventions, and (3) reducing the computational time required for intervention calculations. 
To overcome these challenges, the application of control theory is required; however, traditional control theory has limitations in handling nonlinear, high-dimensional, and complex models such as weather systems \cite{soldatenko2021}.

The problem of identifying an effective intervention reduces to the optimization of its parameters such as timing, location, and intensity. However, the optimization process is extremely challenging due to the vast and complex nature of meteorological phenomena~\cite{bennett1996}. First, obtaining accurate gradient information of the objective function is difficult. Moreover, state-of-the-art numerical weather prediction (NWP) models demand enormous computational resources for each simulation run, thereby enforcing the necessity of minimizing function evaluations during parameter optimization.
These constraints collectively suggest that black-box optimization methods~\cite{Jones1998}, which operate exclusively on input-output data, offer a promising approach for designing interventions. Specifically, such methods iteratively explore the parameter space to seek optimal input values that maximize or minimize the objective function without recourse to derivative information. Their capacity to leverage evaluation results for efficient exploration renders them particularly well-suited to the demands of weather intervention optimization. However, to the best of our knowledge, no previous studies have applied black-box optimization techniques to the design of weather interventions, leaving the question of the most effective algorithm unresolved.

Therefore, in this paper, we design a weather intervention computation method using black-box optimization and evaluate its effectiveness through simulations using NWP models. We formulate two control scenarios: one that permits intervention only at a single time point (initial value intervention) and another that enables sequential interventions via model predictive control.  
Furthermore, by combining these control problems with two experimental settings that differ in scale and complexity, we conduct a comprehensive evaluation of several representative black-box optimization methods. Specifically, we compare the performance of Bayesian optimization, random search, particle swarm optimization, and genetic algorithms.

The contributions of this work are summarized as follows.
This study proposes a black-box optimization framework for weather intervention optimization and demonstrates that effective interventions can be identified with minimal function evaluations. Furthermore, a rigorous comparative analysis of four representative black-box optimization algorithms is performed in terms of total rainfall reduction, with a detailed examination of their operational characteristics. Notably, the experimental results provide evidence that Bayesian optimization achieves superior performance, particularly in high-dimensional search spaces. Overall, the findings support Bayesian optimization as a highly effective and reliable approach for computing weather interventions, outperforming alternative algorithms.

\section{Control Problem Formulation}
This section first describes the meteorological model used in this study, Scalable Computing for Advanced Library and Environment Regional Model (SCALE-RM).
Next, two weather control problems are formulated using this model and the challenges associated with solving the corresponding optimization problems are discussed.
\subsection{SCALE-RM}
SCALE-RM is a NWP model specifically developed for climate research. This model is part of the SCALE software library, which supports weather forecasting across various computational platforms \cite{sato2015,nishizawa2015}.
Due to its versatility and reliability, SCALE-RM has been widely utilized in studies on weather forecasting and atmospheric science \cite{honda2022}.
In weather control research, SCALE-RM plays a vital role in modeling and assessing control methods.This model enables detailed simulations of atmospheric interactions and external interventions, facilitating precise evaluations of control method effectiveness.

In this study, numerical simulations were conducted using two experimental setups provided by SCALE-RM. The first one, the warm bubble experiment \cite{zhang2004impacts}, employs a two-dimensional model to idealize and simulate convective clouds. The second one, the real atmosphere experiment, uses a three-dimensional model to replicate more realistic atmospheric behavior.
An overview of these two experimental settings is provided in Table~\ref{table:WBRA}, and the details of each experiment are described in Sections 4 and 5, respectively.
\begin{table}[tb]
\centering
\caption{Overview of warm bubble experiment and real atmosphere experiment.}
\label{table:WBRA}
\begin{tabular}{@{}p{2.5cm}p{4.5cm}p{4.5cm}@{}}
  \toprule
  \multicolumn{1}{c}{} & \multicolumn{1}{c}{\textbf{Warm bubble experiment}} & \multicolumn{1}{c}{\textbf{Real atmosphere experiment}} \\ \midrule
  \textbf{Objective}  & Ideal settings to reproduce cumulus convection and localized phenomena. & Reproduce large-scale meteorological phenomena using real atmospheric conditions. \\ \midrule
  \textbf{Computational domain} &  Small (e.g., $\SI{10}{km^2}$) & Large (e.g., $\SI{3240000}{km^2}$) \\
  \bottomrule
\end{tabular}
\end{table}

% 先述の通り，SCALE-RMは大気物理を記述する方程式群を数値的に解くことで，大気の状態をシミュレーションする．これらの方程式自体は決定論的であり，初期条件や境界条件が完全に与えられれば，モデルの出力は一意に定まる．
Let $w_t$ denote the atmospheric state variables at a given time $t$, such as potential temperature or humidity, and $f_0$ denote the model representing atmospheric state changes. 
The weather system can be modeled as
\begin{equation}
w_{t+1}=f_0(w_t),\label{eq:basic_system}
\end{equation}
where, $w_t$ is a high-dimensional vector that encapsulates all meteorological state variables distributed in space, and $f_0$ represents an idealized model assuming no noise.
In both experimental setups, the dimensionality of $w_t$ exceeds tens of thousands. 
The model $f_0$  exhibits nonlinearity due to the complexity of processes governing meteorological phenomena, such as turbulence and cloud microphysics.
%乱流過程と雲微物理
% \newcommand{\subsectionspacing}{-3mm}
% \vspace{\subsectionspacing}
\subsection{Weather Control Problem}
% \vspace{\subsectionspacing}
In this study, interventions to the initial atmospheric state in a numerical weather prediction model are considered as control inputs for weather modification, without assuming specific intervention methods. This approach is based on the framework proposed by Ohtsuka et al.~\cite{ohtsuka2024}.
Let $u_t$ denote the intervention at a given time $t$, and $f$ denote the model representing atmospheric state change with the intervention. 
The control system in this study can be modeled as
\begin{gather}
w_{t+1} = f(w_t, u_t). \label{eq:intervention_system}
\end{gather}

SCALE-RM includes variables MOMX and MOMY, which represent atmospheric momentum $\SI{}{[kg \cdot m/s]}$.
In this study, these variables are modified as an intervention in the atmospheric state, corresponding to the manipulation of the wind velocity field.
Furthermore, The objective of the intervention is set to minimization of the total precipitation intensity, expressed in \SI{}{[kg/m^2/s]}, over a specified surface region from $t=1$ \SI{}{[s]} to $t=T_e$ \SI{}{[s]}. NWP models simulate physical phenomena within a spatial framework called the computational domain, which is divided into basic units called grid cells. 
Let $\mathrm{PREC}(t, x, y)$ denote the precipitation intensity at grid cell $(x, y, 0)$ at time $t$.
Then, define $G$ as the set of grid cells where total precipitation is to be minimized.

In this study, we formulate two control problems: one based on initial value intervention, which applies intervention at a single time step, and another using model predictive control, which enables sequential interventions. The following sections present an overview of each control problem and describe the corresponding optimization formulation.

\vspace{-4mm}
\subsubsection{Initial value intervention}
In this control problem, intervention is applied only at the initial time.
Let the intervention to MOMX and MOMY be denoted as $d_X$ and $d_Y$, respectively. The intervention is applied to a single grid cell, with the intervention location given by $(x, y, z)$. Then, this intervention is represented as 
\begin{gather}
u_0 = (d_X, d_Y, x, y, z). \label{eq:intervention_Init}
\end{gather}
Let $U$ denote the set of grid cells $(x, y, z)$ where interventions can be applied.
The optimization problem to be solved can then be formulated as
\begin{align}
     \underset{u_0}{\text{minimize}} \quad & \sum_{t=1}^{T_{e}} \sum_{(x, y)\in G}\mathrm{PREC}(t, x, y)\notag \\
    \text{subject to} \quad & w_{t+1} = f_0(w_t) , \quad t = 1, 2, \dots, T_e-1 \label{eq:Initial_basic_formula}\\
    & w_1 = f(w_0, u_0)\notag \\ 
    %& u_0=(d_X, d_Y, x, y, z), \quad (x,y,z)\in U\\
    & d_X \in [\underline{d}, \bar{d}], \quad d_Y \in [\underline{d}, \bar{d}], \quad (x,y,z)\in U\notag
\end{align}
where $\underline{d}, \bar{d}$ are the boundary values for the amount of change applied to the atmospheric momentum.

\vspace{-5mm}
\subsubsection{Model Predictive Control}
MPC \cite{mayne2000} is a control method that optimizes control inputs while predicting the future state of the control target at each time step.
MPC measures the system output in real time via feedback control and sequentially computes the appropriate control inputs. Due to this characteristic, MPC is considered highly effective, especially for control systems exhibiting chaos and uncertainty, such as meteorological systems.

In this control problem, intervention is applied at multiple time steps.
Specifically, we consider a setting in which interventions are applied every $T_{\text{step}}$ \SI{}{[s]} starting from time $t=0$ \SI{}{[s]}. Here, we define the interval between interventions as one time step. Let $d_{\tau, X}$ and $d_{\tau, Y}$ denote the intervention to MOMX and MOMY at $\tau$ steps ahead of the current time, respectively. The intervention location is represented as $(x_\tau, y_\tau, z_\tau)$, and the intervention at step $\tau$ is then represented as
\begin{gather}
u_{\tau} = (d_{\tau, X}, d_{\tau, Y}, x_\tau, y_\tau, z_\tau). \label{eq:intervention_MPC}
\end{gather}
The prediction horizon is set to $T_f$ steps, and the optimization problem at time $t$ is formulated as
\begin{equation}\label{eq:MPC_basic_formula}
\begin{aligned}
    \underset{v_0, v_1, \dots, v_{T_f-1}}{\text{minimize}} \quad & \sum_{k=1}^{T_{\text{step}} T_f} \sum_{(x, y) \in G} \mathrm{PREC}(t+k, x, y) \\
    \text{subject to} \quad & w_{T_{\text{step}}\tau + 1} = f(w_{T_{\text{step}}\tau}, v_\tau), \quad \tau = 0, 1, \dots, T_f - 1, \\
    & w_{T_{\text{step}}\tau + l + 1} = f_0(w_{T_{\text{step}}\tau + l}), \quad l = 1, 2, \dots, T_{\text{step}} - 1, \\
    %& v_{\tau} = (d_{\tau, X}, d_{\tau, Y}, x_\tau, y_{\tau}, z_{\tau}), \quad (x_\tau, y_{\tau}, z_{\tau}) \in U, \\
    & d_{\tau, X} \in [\underline{d}, \bar{d}], \quad d_{\tau, Y} \in [\underline{d}, \bar{d}], \quad (x_\tau, y_{\tau}, z_{\tau}) \in U.
\end{aligned}
\end{equation}

Among the intervention sequence $v_0, v_1, \dots, v_{T_f-1}$ obtained by solving Optimization Problem  (\ref{eq:MPC_basic_formula}), $v_0$ is applied as the actual intervention applied at time $t$. By executing this operation every $T_{step}$ from $t = 0$ to $T_e$, the accumulated precipitation in the target region $G$ from $t = 1$ to $T_e$, which is the control objective, is reduced.

\subsection{Challenges of Optimizing an Intervention}
This section discusses the challenges and difficulties associated with solving equations \ref{eq:Initial_basic_formula} and \ref{eq:MPC_basic_formula}.
First, atmospheric phenomena are inherently nonlinear and highly sensitive to initial conditions, causing even slight differences to amplify exponentially \cite{lorenz1969}. As a result, the objective function often has a highly complex landscape with non-convexity and numerous local optima. As many optimization methods assume a smooth objective function, convergence becomes challenging.
Second, numerical weather prediction models are computationally expensive, and incorporating high-resolution models like SCALE-RM into the optimization loop significantly increases simulation costs. 
Furthermore, SCALE-RM solves atmospheric equations through numerical time integration. However, due to the complexity of this process, obtaining an explicit analytical gradient with respect to state variables is challenging, making gradient-based methods inapplicable.
Additionally, constraints on momentum modifications $d_X, d_Y$, grid cell selection, and model uncertainties further complicate finding an exact optimal solution.

Given these challenges, finding exact solutions under limited computational resources is impractical. Instead, approximate methods, heuristics, and local optimization techniques are required. 
To address the difficulty of obtaining gradient information directly, this study employs black-box optimization methods.

\section{Evaluation Setting}
This section first provides an overview of black-box optimization methods, followed by a description of the simulation setting used to evaluate their effectiveness.
\subsection{Black-box optimization}
Black-box optimization methods are used to search for optimal input values that maximize or minimize an objective function when the function itself and its constraints are not explicitly given, and only input-output data is available.
These methods are widely applied in various fields, including engineering design, hyperparameter tuning of machine learning models, and simulations of complex systems \cite{Audet2014}.
In such optimization problems, the objective function is often nonlinear, discontinuous, noisy, or computationally expensive. Therefore, selecting an efficient search method is crucial.
Representative black-box optimization methods include Bayesian Optimization (BO), Random Search (RS), Particle Swarm Optimization (PSO), and Genetic Algorithm (GA).
We omit the detailed explanation of the methods due to page limitations.

\subsection{Simulation Setting}
First, we describe the implementation details and hyperparameter settings for each optimization method used in this study.
BO was implemented using the \verb|gp_minimize| function or the \verb|Optimizer| class from the Scikit-Optimize library, adopting the default hyperparameter settings.
For PSO, we employed the Linearly Decreasing Inertia Weight Method (LDIWM) \cite{shi2004particle}, a widely used parameter adjustment technique. PSO requires defining the hyperparameters $c_1$, $c_2$, and $w$, which determine particle movement.
In LDIWM, these hyperparameters are set as $c_1 = c_2 = 2.0$, and $w$ is linearly decreased as follows: $w=w_{\max}-\frac{n(w_{\max}-w_{\min})}{n_{\max}}$ where $n$ is the current iteration count, and $n_{\max}$ is the maximum number of iterations. Following \cite{shi2004particle}, we set $w_{\max} = 0.9$ and $w_{\min} = 0.4$.
For GA, we used real-valued encoding. The selection, crossover, and mutation methods were set as follows: tournament selection, blend crossover \cite{ESHELMAN1993187}, and random replacement with real values. The corresponding hyperparameters were: a tournament size of 3, a crossover rate of 0.8, a blending factor $\alpha$ of 0.5, and a mutation rate of 0.05.
Since the control problems in this study involve a search space with integer values, when applying PSO or GA, the search space is defined in real numbers, and the obtained parameters are rounded to the nearest integers.

Next, we describe the simulation settings designed to ensure a fair evaluation of the control effectiveness of the black-box optimization methods.
In weather intervention optimization, it is crucial to identify effective interventions with a minimal number of function evaluations. Therefore, simulations were conducted with various maximum function evaluation limits—specifically, 15, 30, 50, 100, 150, 200, 250, and 300. Since both PSO and GA require the specification of a population size, we set the population size based on the evaluation limit: a population size of 5 was used when the maximum function evaluations were 15, and a population size of 10 was employed for all other cases.
Furthermore, to account for the inherent randomness of black-box optimization methods, we conducted simulations using 10 different random seeds, applying each optimization method to each seed. The results were then analyzed and discussed.

\section{Warm Bubble Experiment}
This section presents the simulation setup, results, and discussion in the context of the warm bubble experiment.
\subsection{Experimental Settings}
The warm bubble experiment is a widely used benchmark test for evaluating the performance of NWP models \cite{zhang2004impacts}. This experimental setup replicates atmospheric convection processes under idealized conditions. 
As demonstrated by Ohtsuka et al. \cite{ohtsuka2024}, who validated control methods using this setup, the warm bubble experiment also serves as a benchmark for weather control studies.

\begin{figure}[tb]
 \centering
 \includegraphics[scale=0.42]{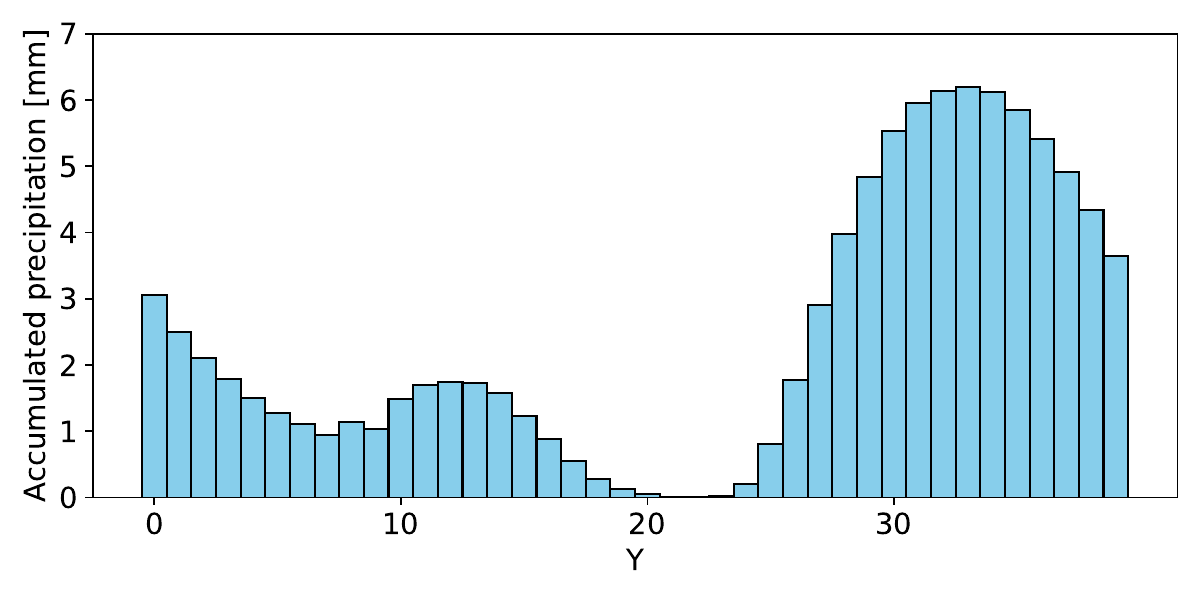}
 \vspace{-5mm}
 \caption{Accumulated precipitation over one hour without any control. \label{fig:WB_CTRL_SumPREC}}
\end{figure}

In this study, simulations were conducted using SCALE-RM version 5.5.1. In this experiment, the horizontal grid spacing was set to $\SI{500}{m}$ in both the east-west and north-south directions, with 1 grid point in the east-west direction and 40 grid points in the north-south direction. The vertical domain consisted of 97 layers, with the model top at $\SI{20}{km}$.
The initial condition featured a warm bubble placed near the surface at the center grid in the $y$-direction. The warm bubble had a horizontal radius of $\SI{4}{km}$, a vertical radius of $\SI{3}{km}$, and a center temperature $\SI{3}{K}$ higher than the surrounding environment. 
This warm bubble triggered cumulus cloud formation and convection, leading to precipitation.
This atmospheric motion is observed over a duration of one hour from the start of the experiment.
Fig.~\ref{fig:WB_CTRL_SumPREC} presents the accumulated precipitation at each location over one hour without intervention.
For each control problem, the objective is to determine an intervention that minimizes the total area of this bar graph, subject to the given constraints.

The two weather control problems described in Section 2  are formulated in a form applicable to the warm bubble experiment.
The control objective is to reduce the total accumulated precipitation over the entire computational domain from time $t=1$ $\SI{}{[s]}$ to $t=3600$ $\SI{}{[s]}$. 
Furthermore, since this experiment employs a north-south vertical plane as a two-dimensional setup, interventions to the atmospheric state are applied only to MOMY.
In this experiment, the lower and upper bounds of $d_Y$, denoted by $\underline{d}$ and $\bar{d}$, were set to $-30$ and $30$, respectively.
Intervention requires predefined intensity limits. However, since this study does not assume a specific intervention method, the appropriate bounds are not uniquely determined. Therefore, based on preliminary experiments, we selected bounds sufficient for evaluating the control performance of each method.

\vspace{-5mm}
\subsubsection{Initial value intervention}
% \textbf{Initial value intervention:}
In this control problem, we attempt to reduce accumulated precipitation by applying intervention only at the initial time step. At the initial time, the intervention applied to MOMY at a specific grid cell $(x, y, z) = (0, y, z)$ is defined as $u_0 = (d_Y, y, z)$. The corresponding optimization problem is then formulated as follows:
% \begin{equation}
\begin{align}
    \underset{u_0}{\text{minimize}} \quad & \sum_{t=1}^{3600} \sum_{y=0}^{39}\mathrm{PREC}(t, 0, y)\notag\\
    \text{subject to} \quad & w_{t+1} = f_0(w_t) , \quad t = 1, 2, \dots, 3599 \notag\\
    & w_1 = f(w_0, u_0), \label{Initial_WB_formula}\\ 
    % \quad & u_0=(d_Y,y, z)\\
    & 0 \leq y \leq 39 , \quad y \in \mathbb{Z}, \notag \\
    & 0 \leq z \leq 96 , \quad z \in \mathbb{Z}, \notag \\
    & -30 \leq d_Y \leq 30. \notag
\end{align}
% \end{equation}

\vspace{-5mm}
\subsubsection{MPC}
In this control problem, we attempt to reduce accumulated precipitation through sequential interventions applied every $T_{\text{step}} = 600$ seconds from the initial time. 
This interval balances the number of control patterns and computational efficiency, allowing effective evaluation of black-box optimization methods.
Let the intervention applied at a future location $(x, y, z) = (0, y_\tau, z_\tau)$ at $\tau$ steps ahead of the current time be represented as $v_\tau = (d_{\tau, Y}, y_\tau, z_\tau)$. With a prediction horizon of $T_f$ steps, the MPC optimization problem at a given time $t$ can be formulated as follows:
\begin{align}
    \underset{v_0, v_1, \dots, v_{T_f-1}}{\text{minimize}} \quad & \sum_{k=0}^{600T_f}\sum_{y=0}^{39}\mathrm{PREC}(t+k, 0, y) \notag\\
    \text{subject to} \quad & w_{600\tau + 1} = f(w_{T_{\text{step}}\tau}, v_\tau), \quad \tau = 0, 1, \dots, T_f - 1,\label{eq:MPC_WB_formula} \\
    & w_{600\tau + l + 1} = f_0(w_{600\tau + l}), \quad l = 1, 2, \dots, 599, \notag\\
    & 0 \leq y_{\tau} \leq 39 , \quad y_{\tau} \in \mathbb{Z}, \notag\\
    & 0 \leq z_{\tau} \leq 96 , \quad z_{\tau} \in \mathbb{Z}, \notag\\
    & -30\leq d_{\tau, Y} \leq 30. \notag
\end{align}

In this experiment, the prediction horizon at the initial time step is set to $T_f = 6$ and decreases by one step as control progresses.
From the sequence of interventions $v_0, v_1, \dots, v_{T_f-1}$ obtained by solving the optimization problem (\ref{eq:MPC_WB_formula}), only $v_0$ is applied as the actual intervention at time $t$.
Since the total experiment duration is 3600 seconds, the optimization problem (\ref{eq:MPC_WB_formula}) is solved a total of six times.

\subsection{Results and Discussion}
\begin{figure*}[!tb]
    \centering
    \begin{minipage}[tb]{0.61\textwidth}
        \centering
        \includegraphics[width=\linewidth]{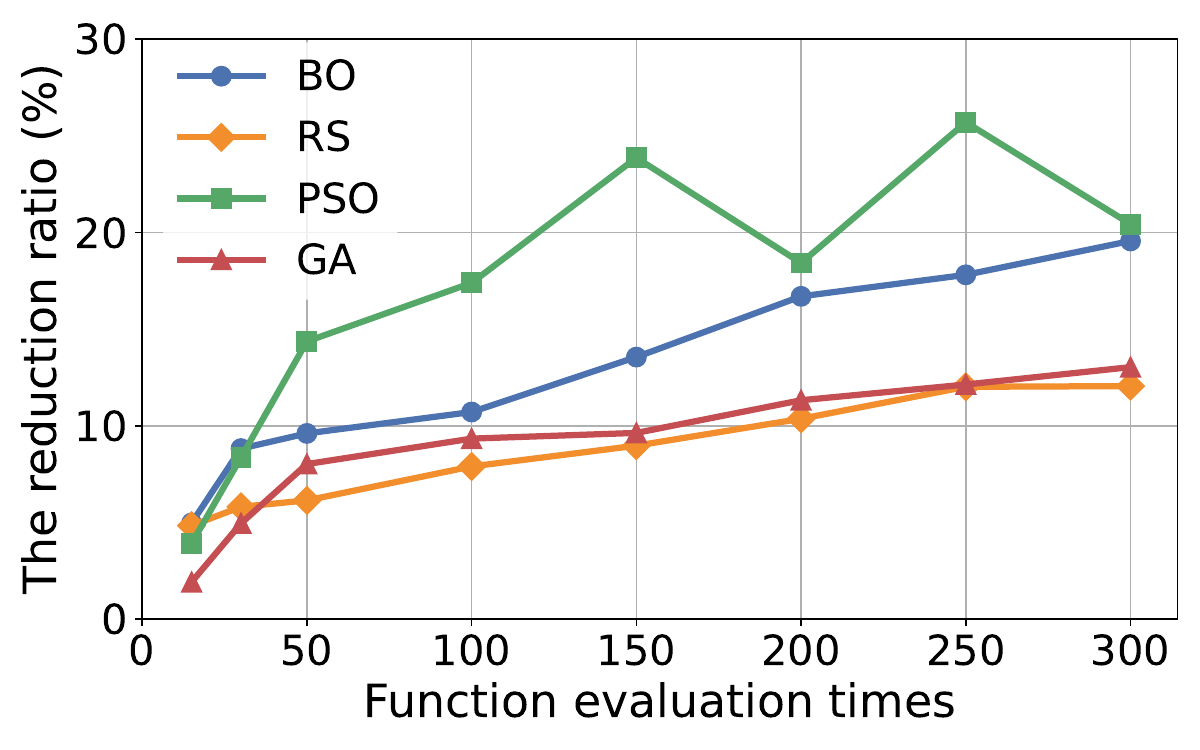}
        % \vspace{-0.3cm}
        \subcaption{}
    \end{minipage}
    \hspace{1.5mm} 
    \begin{minipage}[tb]{0.362\textwidth}
        \centering
        \includegraphics[width=\linewidth]{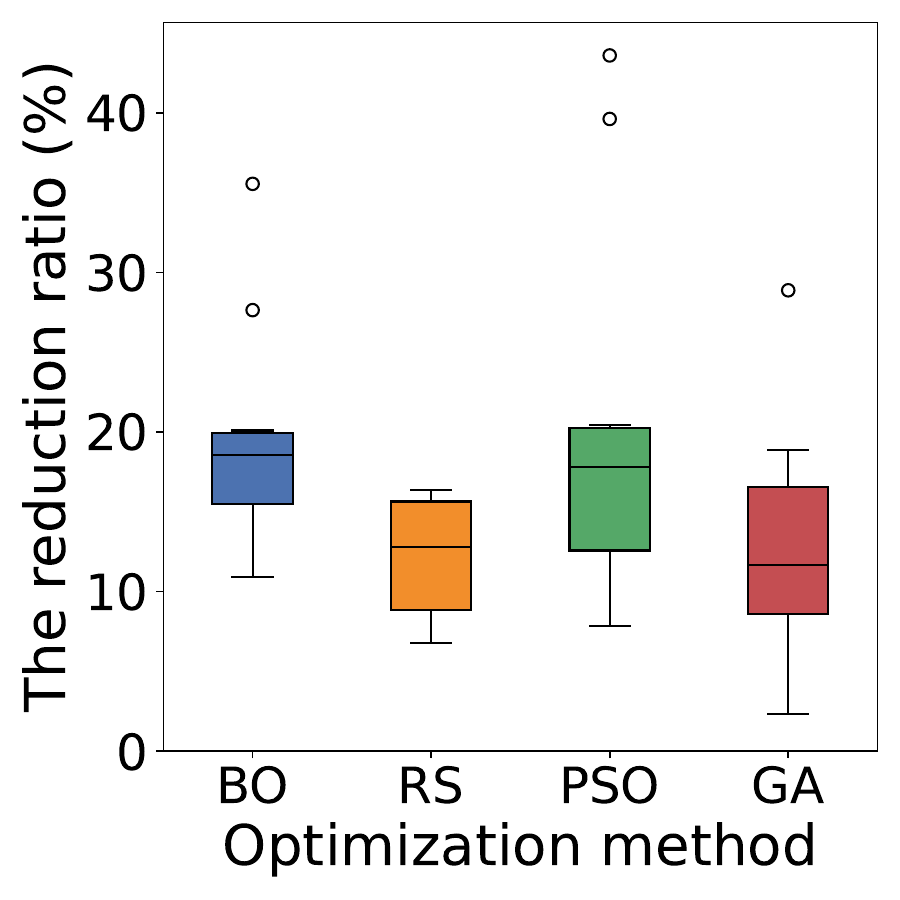}
        % \vspace{-0.1cm}
        \subcaption{}
    \end{minipage}

    \vspace{0.1cm} 
    \begin{minipage}[tb]{0.61\textwidth}
        \centering
        \includegraphics[width=\linewidth]{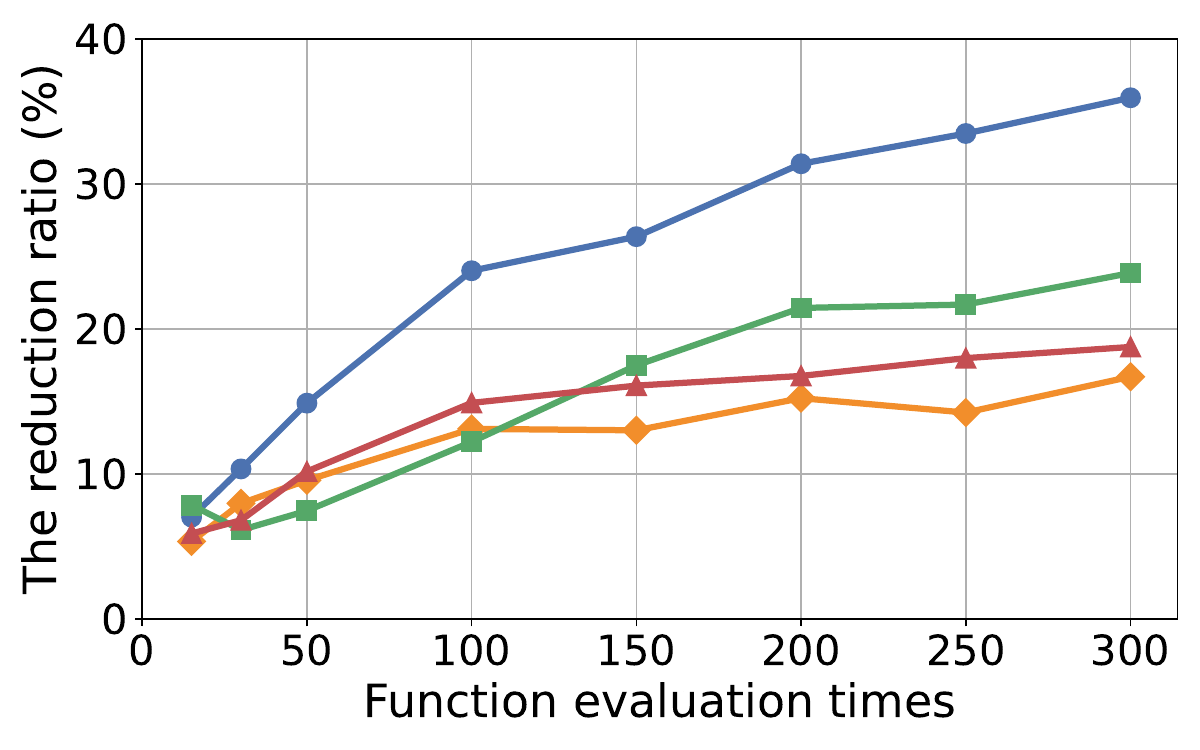}    
        \subcaption{}
    \end{minipage}
    \hspace{1.5mm}
    \begin{minipage}[tb]{0.36\textwidth}
        \centering
\includegraphics[width=\linewidth]{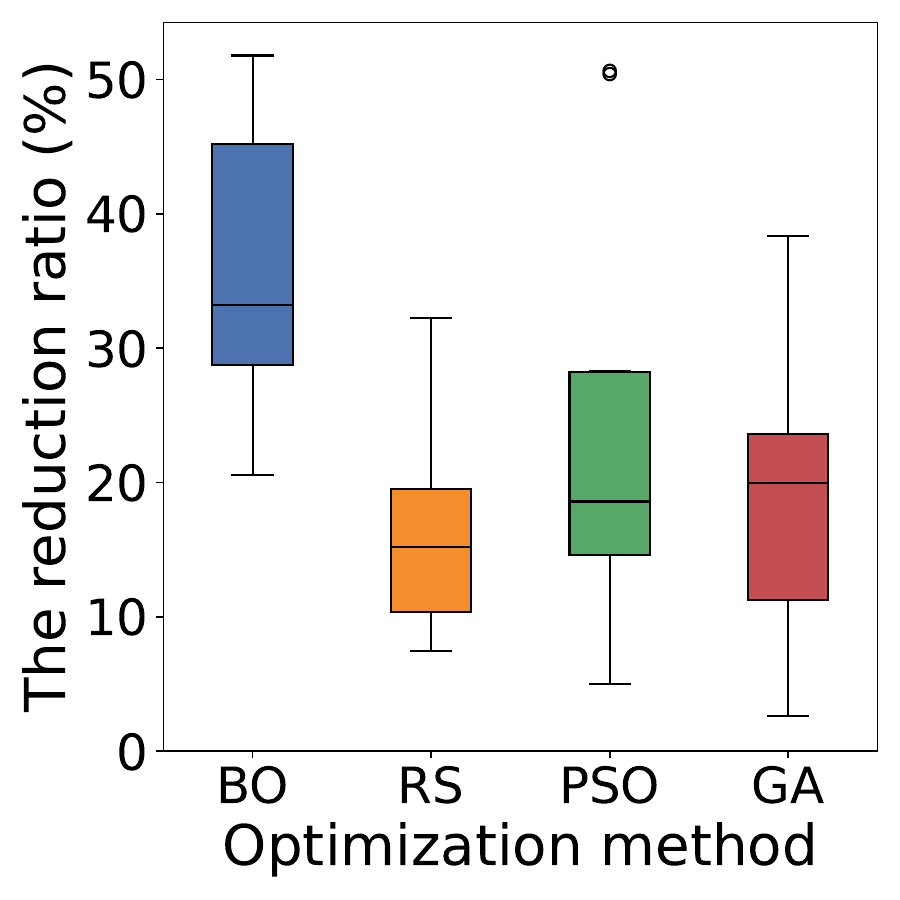}
        \subcaption{}
    \end{minipage}
    \vspace{-2mm}
    \caption{Results of the warm bubble experiment. The upper panel shows the results of initial value intervention, while the lower panel shows those of MPC. (a) and (c) represent the average control effect at each function evaluation step, while (b) and (d) present the results with a function evaluation limit of 300.}
    \label{fig:WB-result}
    % \vspace{-20mm}
\end{figure*}

\begin{figure}[tb]
 \centering
\includegraphics[width=\linewidth]{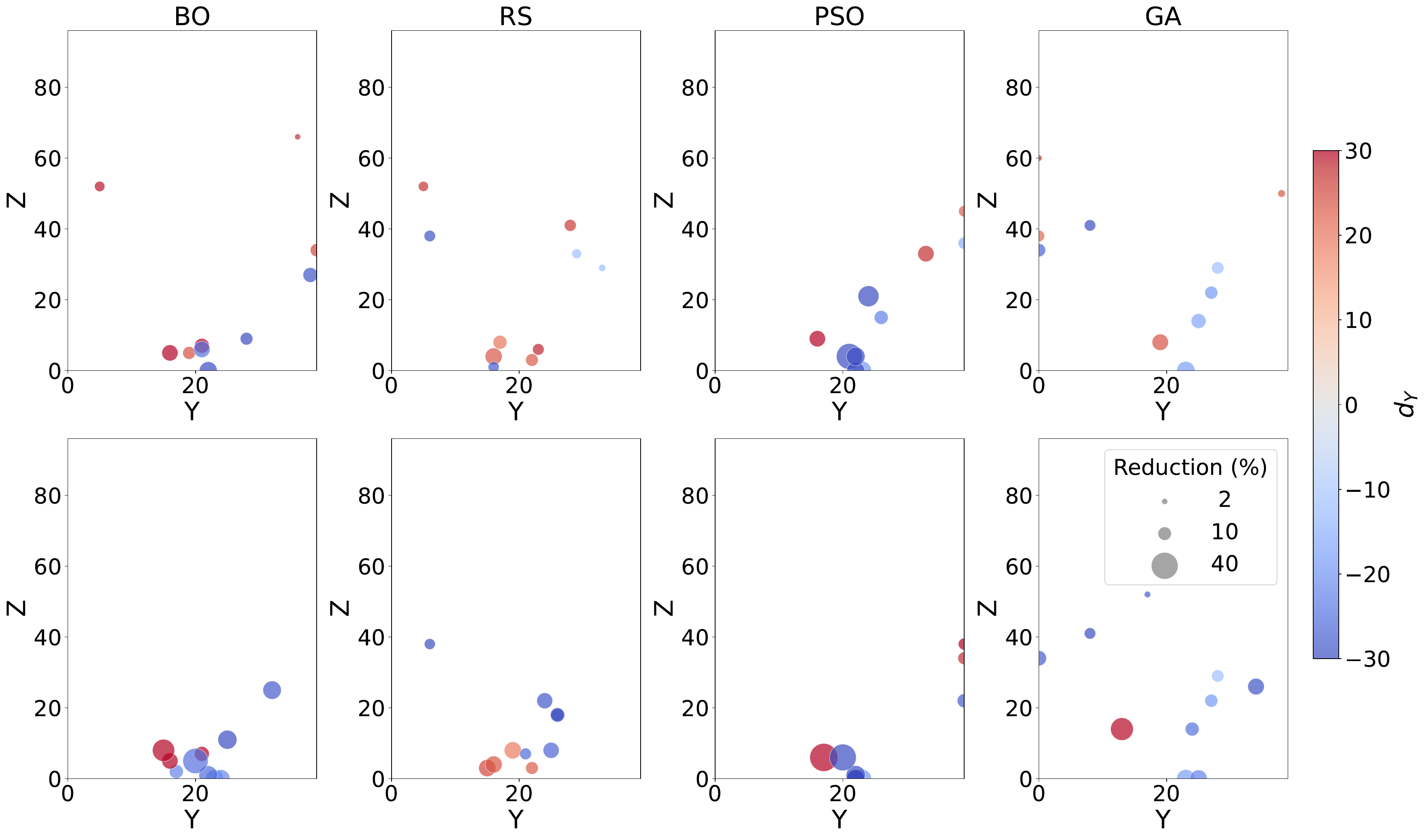}
\vspace{-6mm}
 \caption{The optimal solutions and  corresponding values were obtained for each maximum number of function evaluations in the warm bubble experiment. The optimal solutions obtained from each simulation are represented by the position and color of the points. The size of each point indicates the reduction rate of accumulated precipitation achieved through control. The upper panel shows results for a maximum of 100 function evaluations, while the lower panel shows those for 300 function evaluations.  \label{fig:WB_Init_SC}}
\end{figure}

This study aims to compare the effectiveness of black-box optimization methods as computational approaches for weather intervention.
Since simulations require substantial computational resources, identifying appropriate parameters with a minimal number of function evaluations is crucial.
To this end, for each of the two control problems, the control effectiveness of the optimal solutions obtained under different function evaluation limits is computed and presented in Figs.~\ref{fig:WB-result}a and \ref{fig:WB-result}c.
Furthermore, to evaluate the impact of the inherent randomness in black-box optimization methods on control performance, we conducted 10 simulations with the number of function evaluations fixed at an upper limit of 300.
The results of these simulations are presented in Figs.~\ref{fig:WB-result}b and \ref{fig:WB-result}d.
Finally, the optimal solutions and corresponding optimal values explored by the black-box optimization methods for the control problem involving initial value interventions are presented in Fig.~\ref{fig:WB_Init_SC}.

First, as shown in Figs.~\ref{fig:WB-result}a and \ref{fig:WB-result}c, the reduction rate of accumulated precipitation generally improved with an increasing number of function evaluations, except when PSO was applied to the initial value intervention problem.
For the initial value intervention problem, PSO achieved higher reduction rates, whereas for the MPC problem, BO performed better.
This suggests that differences in control problems, such as the dimensionality of the search space and the use of feedback control, significantly impact control effectiveness.

Next, as shown in Figs.~\ref{fig:WB-result}b and \ref{fig:WB-result}d, BO consistently achieved the highest reduction rate among all methods, even in the worst-case scenarios. This trend was particularly pronounced in the MPC problem, where, with the maximum number of function evaluations fixed at 300, the worst-case reduction rate achieved by BO exceeded the median reduction rates of other methods. 
This result can be attributed to BO’s use of Gaussian process regression, which allows the regression model to approximate the objective function and effectively balance exploration and exploitation. Consequently, the random variability of initial sampling points had minimal impact on the final control performance.
For PSO, outliers with exceptionally high reduction rates were observed in both control problems. This is likely due to PSO’s convergence characteristics—once an exceptionally favorable search point is found, the swarm rapidly converges toward that search point. This suggests that PSO’s control performance is strongly influenced by its inherent randomness.

Finally, Fig.~\ref{fig:WB_Init_SC} shows that grid cells with high control effectiveness, represented by large points, are concentrated in the region where $Z \leq 20$.
Additionally, control effectiveness tends to increase as  $|d_Y|$ becomes larger.
\section{Real Atmosphere Experiment}
This section presents the simulation setup, results, and discussion in the context of the real atmosphere experiment.
\subsection{Experimental Settings}
The real atmosphere experiment is a numerical simulation framework that replicates and predicts real-world weather phenomena based on actual atmospheric conditions and surface parameters.
This framework requires substantial computational resources due to large-scale data processing and complex physical calculations. However, the higher computational cost enables more realistic behavior compared to idealized experiments.
Therefore, this framework serves as a crucial tool for advancing real-world weather control.

In this study, simulations were conducted using version 5.5.1 or 5.5.3 of the SCALE-RM model. In these simulations, the horizontal grid spacing was set to \SI{20}{km} in both the east-west and north-south directions, with 90 grid points in each direction. The model included 36 vertical layers, with an upper boundary approximately \SI{28}{km} above the surface.
The computational domain was centered around approximately \ang{34}N and \ang{135}E, encompassing Japan and the Korean Peninsula.
Atmospheric data, including wind speed and temperature, as well as elevation and sea surface data at 00:00 UTC on 15 July 2007, were used as initial conditions.

\begin{figure}[tb]
 \centering
 \begin{overpic}[scale=0.33]{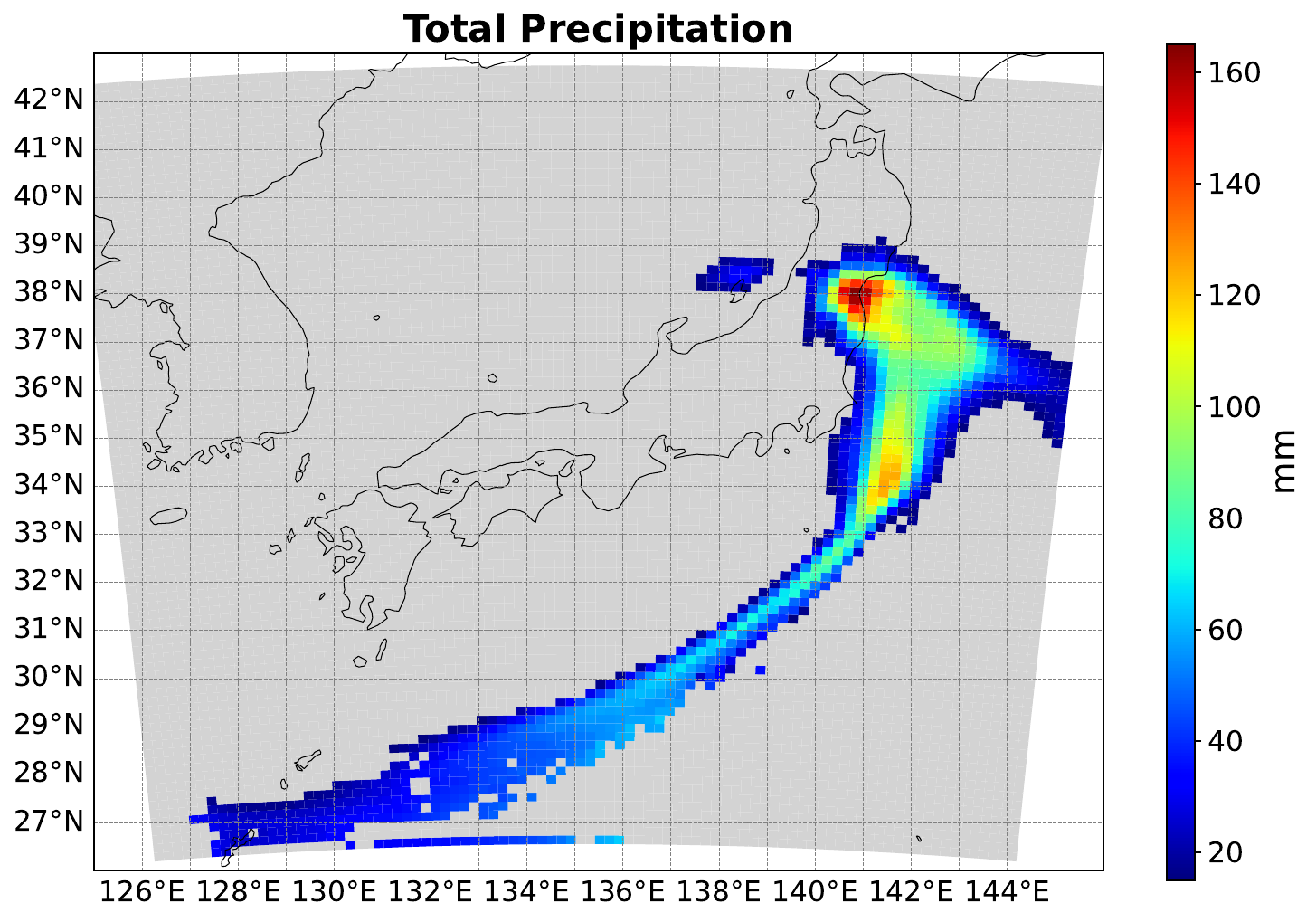}
 \put(62,45){\color{red}\rotatebox{357.5}{\linethickness{1pt}\framebox(8.2,6.5){}}}
 \end{overpic}
 \vspace{-2mm}
 \caption{The 6-hour accumulated precipitation without any interventions is presented, with only grid cells receiving $\SI{20}{mm}$ or more of precipitation being colored. The red-framed region in the figure indicates the target area where the accumulated precipitation is to be minimized in this section.}\label{fig:RA_CTRL_SumPREC}
\end{figure}

The two weather control problems described in Section 2 are formulated in a form applicable to the real atmosphere experiment. 
First, a baseline simulation was conducted without intervention to evaluate the 6-hour accumulated precipitation at each location in the computational domain. The results, shown in Fig.~\ref{fig:RA_CTRL_SumPREC}, indicate significant precipitation within the red-framed region. 
Therefore, the objective of the intervention is set to minimization of the  total precipitation intensity $\SI{}{[kg/m^2/s]}$ over the target region $G=\{(x, y)\in \mathbb{Z}^2 \mid 65\leq x\leq74, 60\leq y \leq 69\}$ during the period from $t=1$ $\SI{}{[s]}$ to $t=21600$ $\SI{}{[s]}$.

In a preliminary experiment, we attempted to control the system using the same settings as those in Section 4, by applying intervention only to a single grid cell. However, the reduction in accumulated precipitation was negligible. 
Here, we expand the number of grid cells to which intervention is applied. 
The set of grid cells where intervention is applied is defined as
\begin{equation} R_{\ell,m,n}=\{(x, y, z)\in \mathbb{Z}^3\mid \ell\leq x \leq \ell+4, m\leq y \leq m+4, n \leq z \leq n+4 \} \label{eq:RA_Control_Domain} \end{equation}
where $\ell, m, n$ represent the western, southern, and bottom boundaries of the region where intervention is applied, respectively. The same intervention is applied to all grid cells within this region.
However, based on the results in Section 4, interventions in the upper atmosphere have limited effects. Therefore, interventions are applied near the surface by fixing $n=0$.
Furthermore, Fig.~\ref{fig:RA_CTRL_SumPREC} indicates that interventions in the western part of the computational domain have a limited impact on precipitation reduction in the target region. 
Based on this observation, the feasible intervention is restricted to $\ell \in \{45, 46, \dots, 85\}$ and $m \in \{0, 1, \dots, 85\}$.

In the simulation, the intervention bounds were initially set to $\underline{d}=-30$ and $\bar{d}=30$, as in the warm bubble experiment. 
However, this sometimes led to a loss of physical consistency, causing simulation failure. To prevent this issue, the bounds were adjusted to $\underline{d}=-20$ and $\bar{d}=20$.

\vspace{-5mm}
\subsubsection{Initial value intervention}
In this control problem, we attempt to reduce accumulated precipitation by allowing intervention only at the initial time. An intervention is applied at the initial time to the region $R_{\ell,m}=\{(x, y, z)\in \mathbb{Z}^3 \mid \ell \leq x \leq \ell +4, m\leq y \leq m+4, 0 \leq z \leq 4 \}$, which is represented as $u_0=(d_X, d_Y, R_{l,m})$.
We omit the details due to page limitation.

\vspace{-5mm}
\subsubsection{MPC}
In this control problem, we attempt to reduce accumulated precipitation by allowing sequential interventions.
The intervention is applied at intervals of $T_{\text{step}}=3600$, starting from the initial time. 
Let the intervention applied to a region $R_{\tau, \ell,m}=\{(x, y, z)\in \mathbb{Z}^3\mid \ell_\tau \leq x \leq \ell_\tau +4, m_\tau \leq y \leq m_\tau +4, 0 \leq z \leq 4 \}$ at $\tau$ steps ahead of the current time be represented as $v_\tau=(d_{X, \tau}, d_{Y, \tau}, R_{\tau, \ell,m})$.
In this experiment, the prediction horizon $T_f$ at the initial time step is set to 6.
We omit the details due to page limitation.

\subsection{Results and Discussion}

\begin{figure*}[!tb]
    \centering
    \begin{minipage}[tb]{0.61\textwidth}
        \centering
        \includegraphics[width=\linewidth]{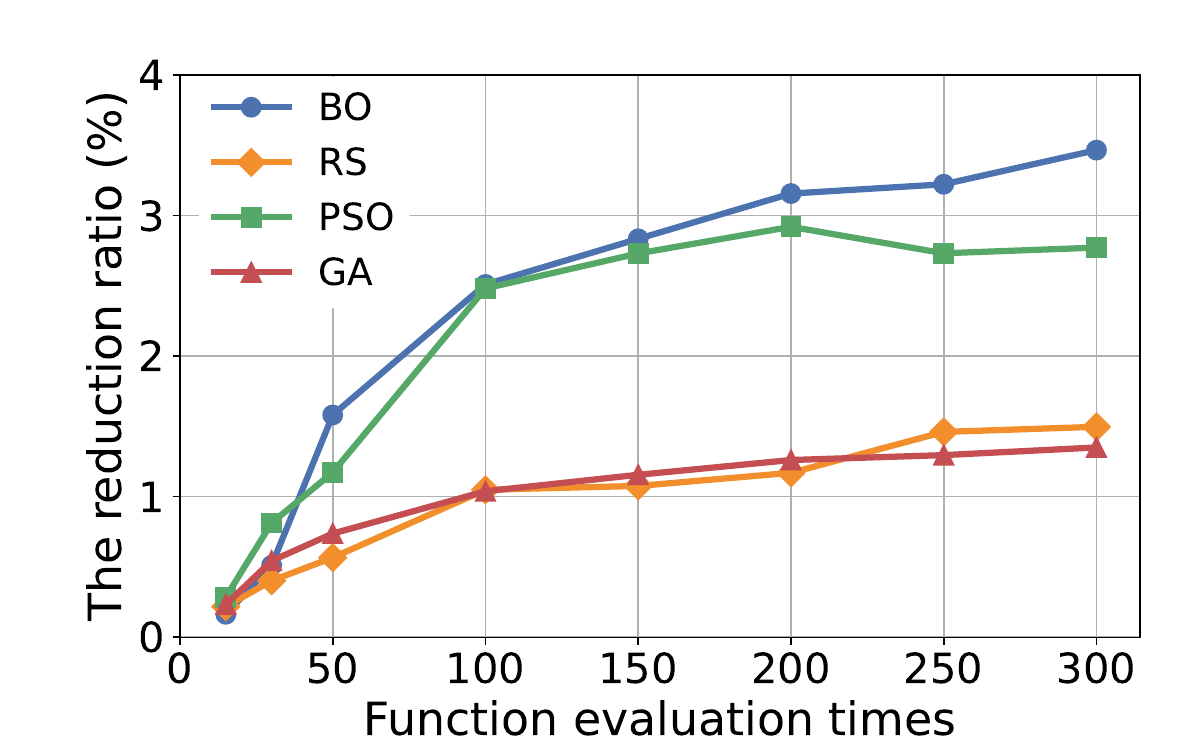}
        \subcaption{}
    \end{minipage}
    \hspace{1.5mm} % 隙間を開けるためのコマンド
    \begin{minipage}[tb]{0.36\textwidth}
        \centering
        \includegraphics[width=\linewidth]{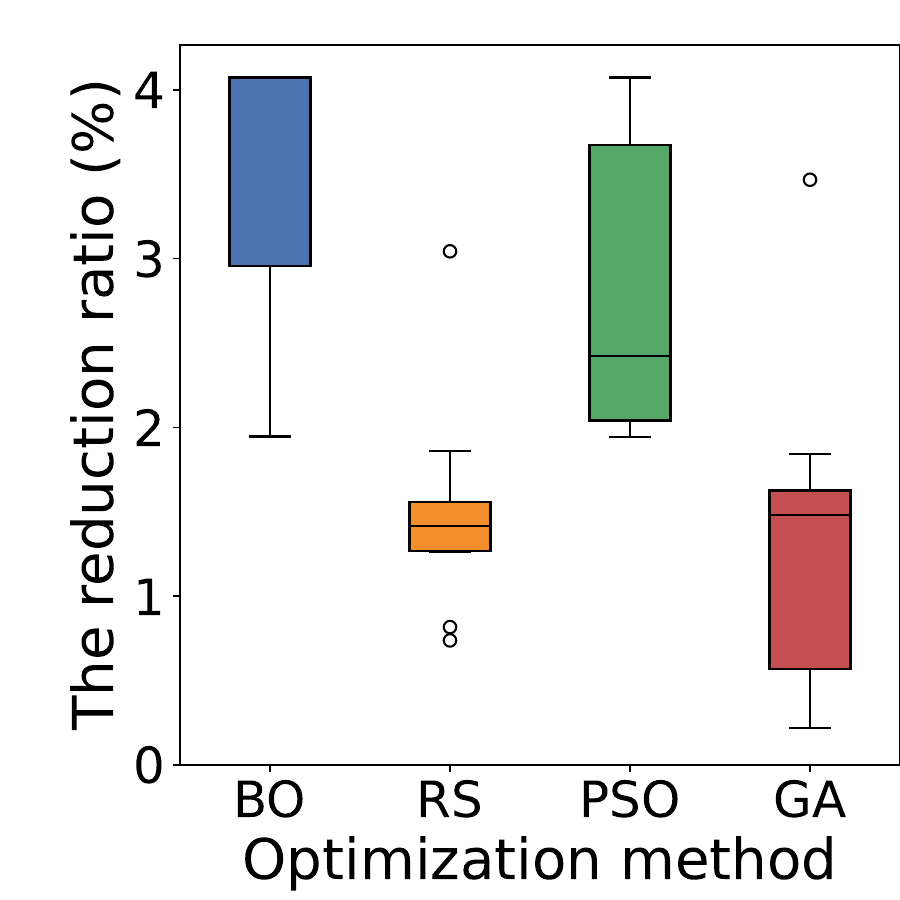}
        \subcaption{}
    \end{minipage}
    % \vspace{0.1cm} % セット間の余白を追加
    \begin{minipage}[tb]{0.61\textwidth}
        \centering
        \includegraphics[width=\linewidth]{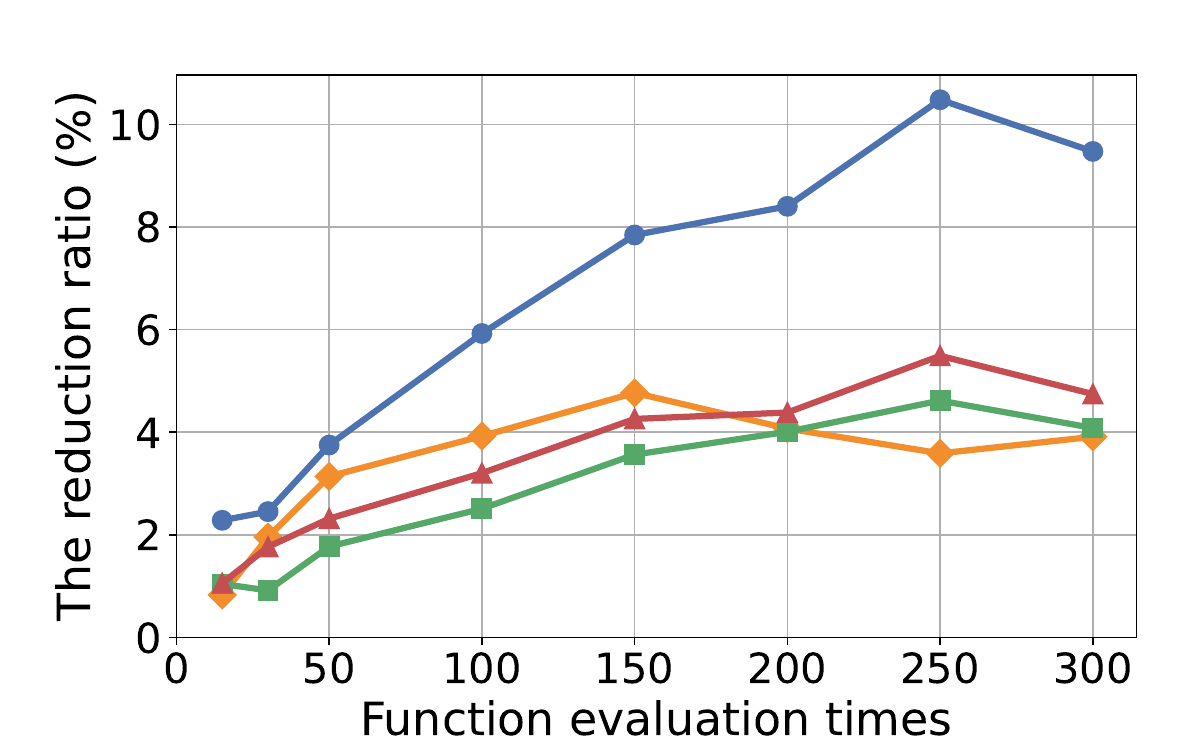}
        \subcaption{}
    \end{minipage}
    \hspace{1.2mm}
    \begin{minipage}[tb]{0.36\textwidth}
        \centering
\includegraphics[width=\linewidth]{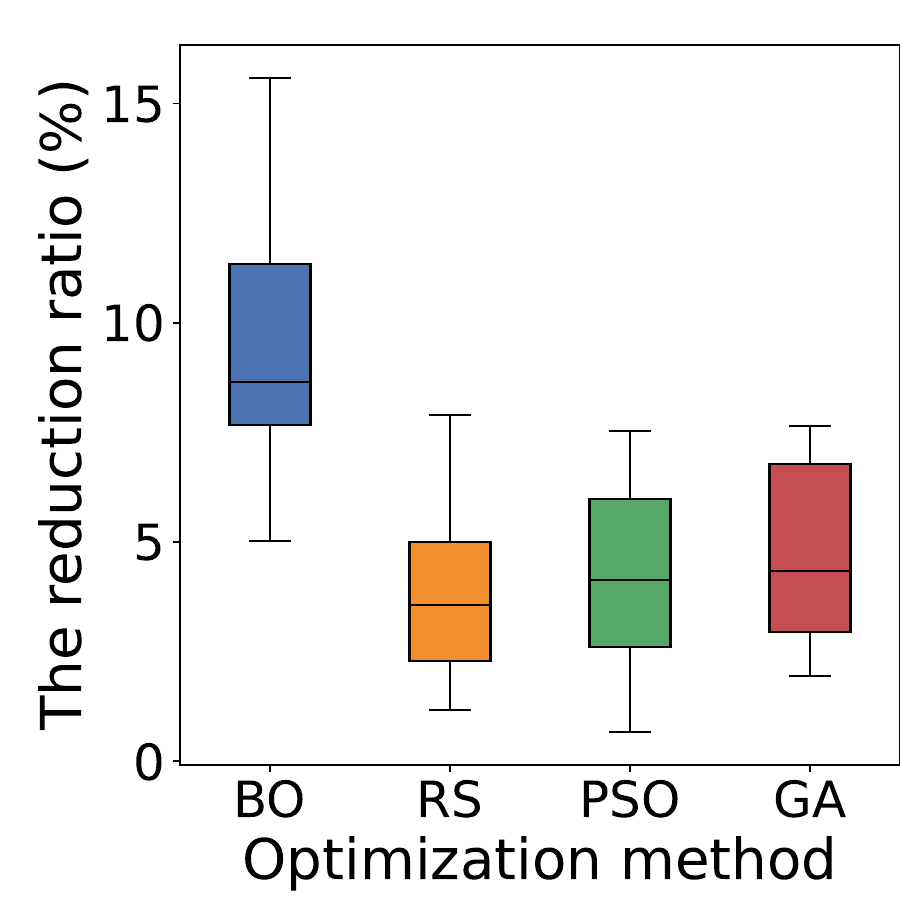}
        \subcaption{}
    \end{minipage}
    \vspace{-2mm}
    \caption{Results of the real atmosphere experiment. The upper panel shows results with initial-value intervention, while the lower panel shows those with MPC. (a) and (c) represent the average control effect at each function evaluation times, while (b) and (d) present results with a function evaluation limit of 300.}
    \label{fig:RA-result}
\end{figure*}
In this section, simulation experiments similar to those in Section 4 are conducted.
The control effect achieved with the optimal solutions was calculated for each upper limit on the number of function evaluations and the results are illustrated in Figs.~\ref{fig:RA-result}a and \ref{fig:RA-result}b. Furthermore, the results of 10 simulations conducted with the number of function evaluations fixed at an upper limit of 300 are presented in Figs.~\ref{fig:RA-result}b and \ref{fig:RA-result}d.

First, as shown in Figs.~\ref{fig:RA-result}a and \ref{fig:RA-result}c, the reduction rate of accumulated precipitation generally improved with an increasing number of function evaluations, except when PSO was applied to the initial value intervention problem, similar to the warm bubble experiment. Furthermore, in both control problems, BO achieved the highest reduction rates among the four methods.
Next, as shown in Figs.~\ref{fig:RA-result}b and \ref{fig:RA-result}d, BO consistently achieved higher minimum reduction rates than other methods, aligning with the results of the warm bubble experiment. Notably, in the initial value intervention problem, when the number of function evaluations was fixed at 300, the median reduction rate equaled the maximum reduction rate, indicating exceptionally high convergence to the optimal solution with BO.
A significant difference from the warm bubble experiment is that BO was also the most effective method for the initial value intervention problem. In the real atmosphere experiment, the search space dimensionality increased by one, which may have influenced the results. This is further supported by the observation that in the MPC problem, which has an even higher-dimensional search space, the control effectiveness of BO becomes even more pronounced.
\section{Conclusion}
In this study, we designed a weather intervention computation method based on black-box optimization and evaluated its effectiveness through simulations using the NWP model. The results suggest that, among the black-box optimization methods tested, BO demonstrates notably high effectiveness. 
It should be noted that these findings are based on specific experimental settings and may not necessarily generalize to other scenarios. Nevertheless, our results indicate the potential of black-box optimization as a viable approach for weather intervention computation and offer valuable insights into the distinct characteristics of each method.

Finally, two future challenges remain.
First, hyperparameter selection was not fully optimized. PSO and GA have hyperparameters that significantly impact search performance. In this study, they were set based on previous research and preliminary experiments, but their suitability for the specific problems addressed remains unverified.
Second, the feasibility of the optimal interventions in real atmospheric conditions is uncertain. This study applied interventions to the wind field over approximately $\SI{10000}{km^2}$, but implementing such large-scale interventions remains a major challenge. 
Designing control problems based on realistic interventions and verifying their feasibility through simulations can contribute to assessing the viability of weather control.

\begin{credits}
\subsubsection{\ackname} 
This work was supported by JST Moonshot R\&D Program Grant Numbers JPMJMS2284.
% A bold run-in heading in small font size at the end of the paper is
% used for general acknowledgments, for example: This study was funded

\subsubsection{\discintname}
The authors have no competing interests to declare that are relevant to the content of this article.
% It is now necessary to declare any competing interests or to specifically
% state that the authors have no competing interests. Please place the
% statement with a bold run-in heading in small font size beneath the
% (optional) acknowledgments\footnote{If EquinOCS, our proceedings submission
% system, is used, then the disclaimer can be provided directly in the system.},
% for example: The authors have no competing interests to declare that are
% relevant to the content of this article. Or: Author A has received research
% grants from Company W. Author B has received a speaker honorarium from
% Company X and owns stock in Company Y. Author C is a member of committee Z.
\end{credits}

\end{document}